# Use of full-scale hybrid horizontal tubular photobioreactors to process agricultural runoff

María Jesús García-Galán[1*], Raquel Gutiérrez[1], Enrica Uggetti[1], Víctor Matamoros[2], Joan García[1], Ivet Ferrer[1]

[1] GEMMA - Environmental Engineering and Microbiology Research Group, Department of Hydraulic, Maritime and Environmental Engineering, UniversitatPolitècnica de Catalunya·BarcelonaTech, c/ Jordi Girona 1-3, Building D1, E-08034, Barcelona, Spain

[2] Department of Environmental Chemistry, IDAEA-CSIC, c/Jordi Girona, 18-26, E-08034, Barcelona, Spain

[*] Corresponding author: María Jesús García-Galán

Tel.: +34 934016204

Fax: +34 934017357

E-mail address: chus.garcia@upc.edu

**Highlights**

- Agricultural runoff was treated in a microalgae-based treatment system

- A full-scale tubular photobioreactor was tested during 4 months
- Initial batch tests proved the high microalgae production capacity of the system
- Despite the low nutrient availability, the system achieved 76 g TSS/$m^3$d in spring
- Removal of relevant emerging contaminants was successfully evaluated

**Graphical Abstract**

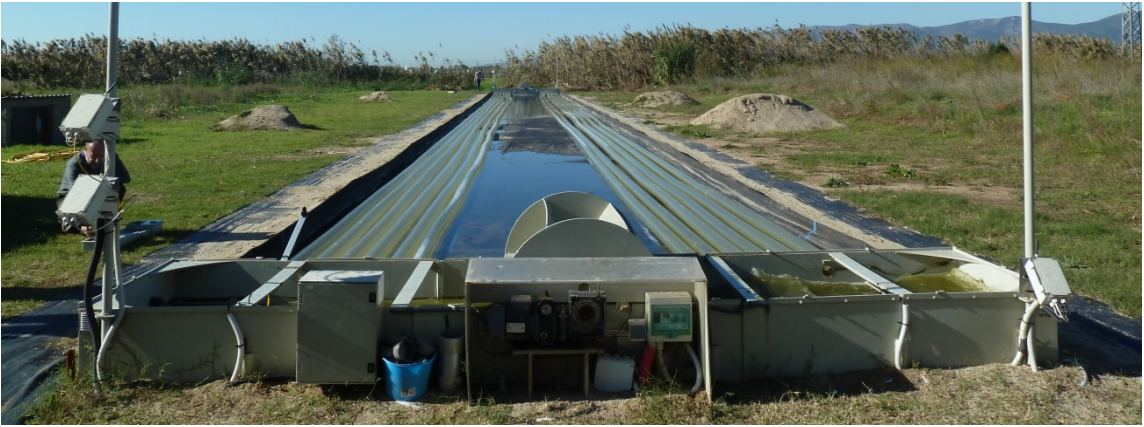

**Abstract**

Diffuse pollution in rural areas due to agricultural run-off is a widespread and difficult problem to address due to the vast areas affected. Drainage channels do receive these polluted waters, but its introduction in the conventional treatment network is unfeasible. Within this context, microalgae-based treatment systems could be used as alternative treatment plants. In this study, a new design of semi-closed (hybrid) tubular horizontal photobioreactor (HTH-PBR) with low energy requirements has been evaluated for microalgae cultivation at full-scale (8.5 $m^3$), using agricultural runoff as feedstock. This novel system was tested in batch and continuous mode during a total of 4 and 135 days. Considering a full-scale application in an agricultural context, a batch test was carried out to evaluate the performance of the system. An increase of 22% in the biomass concentration in 4 days was registered, and all nutrients were consumed during the first two days. In the continuous experiment (December- April), a productivity between 2-14 g TSS/$m^3$d was reached in winter, whereas values up to 76.4 g TSS/$m^3$d were reached at the end of the study in spring, despite the low nutrients concentration in the feedstock. The elimination of emerging contaminants was also evaluated, obtaining the highest removals for the fragrances tonalide and galaxolide (73% and 68%), and the antiinflamatory diclofenac (61%).

## 1. INTRODUCTION

Aquatic ecosystems are constantly receiving a broad range of organic micropollutants from anthropogenic origin, from both point sources such as wastewater treatment plant (WWTP) effluents (urban, rural or industrial) (Verlicchi et al., 2010; Gros et al., 2012; García-Galán et al., 2016), to non-point sources such as urban or agricultural runoff waters after strong rain events (Dolliver et al., 2008; Topp et al., 2008; Sabourin et al., 2009). The equilibrium of river and stream ecosystems can be jeopardized if their capacity to attenuate and neutralize these inputs (dilution, biodegradation, etc) is overcome. Groundwater systems are also indirectly affected if polluted rivers are feeding the aquifers by infiltration or by artificial recharge with surface or reclaimed water (Díaz-Cruz et al., 2008).

Agricultural fields are probably the main source of diffuse pollution to both surface and groundwater systems in rural areas. Inorganic fertilizers and cattle manure (regarded as a very valuable fertilizer containing high amounts of N, P, K, organic carbon (OC), etc) are regularly applied in crop fields during the harvesting season, together with pesticides and herbicides in the pre-harvest season. In consequence, as these compounds are usually highly soluble in order to translocate easily in the soil to reach below ground structures of plants, agricultural runoff can contain a wide variety of contaminants including the excess of herbicides and pesticides, heavy metals and nutrients (N/P). A clear example of contamination by agricultural runoff is the high level of nitrates from fertilizers traditionally found in surface water bodies, leading to algae blooms that consume all the $O_2$ available and don´t let the sunlight reach the benthic photosynthetic organisms in lakes or rivers.

Intensive cattle farming is also another relevant non-point pollution source, since a wide variety of veterinary pharmaceuticals (PhACs), mainly antibiotics, are used in the prevention and treatment of microbial infections. Depending on the drug, livestock can excrete high amounts of the administered dose as metabolites and also as the unchanged parent substance (Dolliver et al., 2008; Tong et al., 2009; Wei et al., 2011). The extensive use of manure from medicated animals in crop fields is considered by different authors as one of the major routes by which veterinary antibiotics enter the environment, as once applied on the top soil, these excreted residues can both percolate and reach groundwater bodies or reach surface waters during storm events (Stoob et al., 2007; Dolliver et al., 2008; Watanabe, et al., 2010; Kwon, 2011). Similarly, biosolids application as soil amendment is also a common practice and an option favoured from governments for sludge management, as it results in the recycling of nutrients and it also improves the soil structure (Topp et al., 2008; Edwards et al., 2009; Sabourin et al., 2009). Several studies have focused in the incomplete removal of organic and emerging contaminants in WWTPs, which are still present in water effluents discharged in river courses, and in the case of the more lipophilic compounds, in digested sewage sludge. For instance musks, UV blockers, flame retardants, bisphenol A or triclosan are frequently detected in biosolids (Plagellat et al., 2006; Clarke et al., 2011; Gago-Ferrero et

al., 2011).

Drainage or open channels can receive an important part of this agricultural runoff, which will finally discharge into rivers or agricultural irrigation channels. Moreover, these irrigation channels usually receive wastewater (treated or raw) as well, especially in non-developed countries leading to an increase in the agricultural production of the area due to the recycling of nutrients in the crop fields (Lees et al., 2016; Christou et al., 2017). However, when these channels discharge directly into a river, potential contamination is spreading and it may affect a large number of non-target species. It would be too costly to direct these drainage channels from agricultural fields into main collectors towards WWTPs. Therefore, local and alternative treatments should be considered.

During the last years, wastewater treatment systems based on microalgae have been investigated with great interest due to the high capability of microalgae to remove nutrients, heavy metals and bacteria (Muñoz et al., 2006; Abdel-Raouf et al., 2012). Indeed, microalgae can grow in low quality water, such as wastewaters, as these still contain high amounts of nutrients (N, P), essential for microalgae production (Pawar, 2016). The use of wastewater as feedstock for microalgae biomass growth leads to a dual benefit, as microalgae have proved to be highly efficient in removing these nutrients, producing clean water as a by-product, and at the same time the production of algal biomass increases; this biomass will be further processed and converted to bioenergy. Furthermore, it has been demonstrated that, under specific growing conditions, different added-value products can also be recovered from microalgae, such as glycogen or bioplastics (Arias et al., 2017). From an environmental perspective, this biological wastewater treatment with microalgae for nutrient removal is considered among the most environmentally favorable and less expensive treatments due to the reduced energy requirements, being ideal for their implantation in rural areas. Lundquist et al. (Lundquist et al., 2010) reported that microalgae production costs would decrease approximately a 10% if wastewater was used as culture media. A uniform exposure of light to microalgae cells would be the main factor to consider in order to maximize both the use of outdoor solar radiation and the absorption of $CO_2$ and to obtain an optimum concentration of microalgae (Vasumathi et al., 2012).

Currently, there are two main systems for microalgae bioremediation (traditionally used for biomass production), open and closed systems. Open systems (e.g. open raceway or high rate algal ponds, HRAPs) have already been used for decades and for different applications (Oswald, 1990), due mainly to their lower costs and energy consumption. Nowadays, nearly all the microalgae biomass produced worldwide (approximately 5000-6000 tons of dry biomass) is obtained from HRAPs (Pulz, 2001). There are other major operational differences between open and closed photobioreactors (PBRs) that can influence the microalgae growth in both systems as summarized in Table S1 of the Supplementary Information (SI). However, open systems are subjected to a poorer control of environmental parameters (i.e. temperature, salinity, and solar radiation) and they are under a higher risk of contamination (predators, fast growing heterotrophs) (Yuan et al., 2009). In contrast, closed PBRs provide a better pH follow up, better protection against culture contamination, better mixing and less evaporative loss (Molina et al.,1999). Yet, closed systems have also some main drawbacks such as higher material and maintenance costs, problems with toxic accumulation of dissolved oxygen (DO), overheating, bio-fouling and the difficulty to scale up (Molina et al., 1999; Mata et al., 2010). Among the different designs of closed PBRs, tubular PBRs are the only type used at large scale (Chisti, 2007).

Regardless of the advantages and disadvantages of both reactors, the combination of open and closed reactors has been proposed as the most effective configuration for growing algae. In this context, the objective of the present work was the design and operation of a full-scale Hybrid Tubular Horizontal Photobioreactor (HTH-PBR) to treat agricultural run-off water in a rural area near Barcelona (Spain). The efficiency of the system was followed up during the first 4 months of operation (15$^{th}$ December-20$^{th}$ April, 135 days), in terms of agricultural runoff remediation and biomass production.

## 2. MATERIALS AND METHODS

### 2.1. HTH-PBR design and operation considerations.

#### 2.1.1. Description of the full-scale HTH-PBR

Considering the advantages of both HRAP and closed tubular PBR configurations, a HTH-PBR was conceived, designed and constructed by the GEMMA Research Group (Universitat Politècnica de Catalunya-BarcelonaTech) in collaboration with Photobioreactores S.L. The reactor consisted of both open and closed units (see Figure S1 in Supplementary Information (SI)):

- The open unit consisted of two tanks (tank E and tank W) made from polypropylene (3.5 m width x 1 m length x 0.4 m height) (Figure 1). Both tanks ensure and favour the homogenous distribution and mixing of the liquor and also to release the exceeding DO, accumulated along the closed tubes. A paddle-wheel with six blades (1 m width x 0.35 m long) was installed 1.8 m away from the external edge in each tank and at 3 cm height from the bottom. An engine (750 W) connected to the paddle wheels provided a turning speed from 0 to 12 rpm. Both tanks were covered with plastic translucent plates, thus only the space occupied by the paddle wheels was exposed to the open air. Each tank was maintained at a working volume of 0.612 m$^3$ (0.17 m depth fixed), which corresponded to the 44% of the volume of each tank and a 14% of the total HTH- PBR capacity.
- The closed unit was constituted by twelve transparent polyethylene tubes of 125 mm diameter and 50 m length connecting the two tanks. The tubes provided a volume of 0.6 m$^3$ each, reaching a total working volume of 7.2 m³ (86% of total volume). The tubes were laid down on a waterproof covering sheet in order to ensure a separation from the ground surface. Besides, in order to refrigerate the tubes, this covering sheet could be filled with cold water covering two-third of their volume. However, this would only be necessary when temperatures exceeded 30º C.

The HTH-PBR was installed in Agrópolis, an experimental area belonging to the UPC facilities (Barcelona, 41.288 N, and 2.043 E UTM). Although the land required for the installation of the whole reactor was 182 m², the working surface area (tubes and tanks) corresponded to only 82.2 m². All the technical characteristics of the HTH-PBR are summarized in Table 1. The following online sensors were installed in order to control the HTH-PBR operation: (1) two temperature sensors (one in each tank), (2) two pH sensors (one in each tank), (3) a DO sensor (in tank E), (4) a solar radiation sensor (placed close to tank E), (5) an electrical conductivity sensor (in tank E), (6) two level sensors (one in each tank) and (7) two rotation sensors turbines (one in each tank). The HTH-PBR system was controlled via a Programmable Logic Controller (PLC) connected to a computer, by means of a supervisory control and data management system (Green web manager 2.0).

### 2.1.2. *Flow direction and distribution*

The mixed liquor flowed by gravity as a result of the different water level (0.08 m of variation) induced by the paddlewheels. As shown in Figure 1, mixed liquor flowed from the side B of tank E (0.21 m of water level) to the side A of tank W (0.13 m) through six tubes and then return (from the side B of tank W to the side A of tank E) through the remaining six. In order to improve the reactor operation, two breakwaters were placed on the bottom of both tanks (side B) to homogenize the mixed liquor and equal the water level. This way, nearly the same flow velocities were reached in all tubes. Afterwards, the power output of the paddle wheel engine was set to reach velocities which ensured turbulent flow inside the tubes. According to the literature, the velocity inside the tubes should range between 30 cm/s and 50 cm/s to prevent biofouling formation on the tube walls and microalgae biomass sedimentation. The minimum velocity to avoid both events should be 15 cm/s (Contreras et al., 2003). Given the geometry of the reactor (0.4 m height of the tanks) and the diameter of the tubes (125 mm), it was not possible to achieve much higher velocities than that. The turning velocity of 8-9 rpm in the paddle wheel was enough to achieve velocities between 15 cm/s and 19 cm/s inside the tubes and to provide the necessary turbulence inside them. Indeed, following Equation 1, such velocities corresponded to a Reynolds number (Re) between 7000 and 8000, well above 4000, which defines the flow variation from transitional to turbulent (Holman, 2002).

$$Re = \frac{\rho * v * D}{\mu} \qquad \text{Eq. 1}$$

where "ρ" is the culture density (1 kg/m$^3$); "v" is the mixed liquor velocity (m/s); D is the tube diameter (0.125 m) and "μ" is the dynamic viscosity of the mixed liquor (0.003 kg/m sec).

### *2.1.3. Drainage channel and agricultural runoff*

The water input for the HTH-PBR was derived from a drainage channel located on the west part of Agrópolis by pumping. The channel runs through different crop fields in the area, gathering different water runoffs, and it also receives the discharge from the urban WWTP in the Gavà-Viladecans area before reaching the Llobregat estuary. This WWTP works at a capacity of 300 000 population equivalents (PEs), with a design to treat a flow of 64 000 m$^3$/day. Main characteristics of the feedstock are summarized in the next sections.

## 2.2. Experimental design

### *2.2.1. Operational conditions*

This study was performed in two different experimental periods, performing both a batch test and a continuous test. Before the start, 0.5 m$^3$ of microalgal biomass were inoculated from an experimental HRAP located in UPC facilities (Matamoros, et al., 2015). The reactor was operated at full volume capacity (12 tubes, 8.5 m$^3$) during the batch and continuous tests (4 days and 5 months, respectively). Throughout the latter, a volume of 0.5 m$^3$ per day was replaced with water from the irrigation canal, corresponding to 16 days of hydraulic retention time (HRT).

### *2.2.2. HTH-PBR performance*

Data for temperature, DO, pH and solar radiation was registered every 5 minutes by means of online sensors. In order to evaluate the system performance, total and volatile solids (TS and VS) and total suspended solids (TSS) were measured in alternate days. Total and soluble chemical oxygen demand (COD and sCOD), ammonia nitrogen (N-NH$_4^+$), nitrates NO$_3^{2-}$-N), phosphates (P-PO$_4^{3-}$) and alkalinity were monitored weekly in the agricultural run-off and the mixed liquor of the PBR. All samples were always taken at the same time (12.00 PM) and the analyses were carried out according to Standard Methods (APHA-AWWA-WPCF, 1999). To evaluate microalgae biomass production, the turbidity of the mixed liquor samples was measured daily with a Hanna Microprocessor Turbidity Meter HI93703 and was evaluated in terms of TSS (mg/L).

*2.2.3 Microalgae characterization*

At the end of the spring period in the continuous experiment, microalgae biomass characterization was carried out by optical microscopy. Qualitative results showed that the main microalgae species belonged to the genus *Pediastrum* sp., *Chlorella* sp., *Scenedesmus* sp., and the cyanobacteria *Gloeothece* sp (Figure S2). Microalgae genus were identified by microscopic analysis using a Zeiss microscope Axioskop 40 and considering conventional taxonomic books (Palmer, 1962; Bourelly, 1966).

*2.3.3. Analysis of emerging contaminants*

During the continuous test, a sampling campaign was conducted to evaluate the removal of emerging contaminants in the PBR. The presence of ten different compounds including pharmaceuticals (ibuprofen, diclofenac, carbamazepine, clofibric acid, caffeine and triclosan), personal care products (galaxolide, tonalide) and other chemicals such as flame retardants, corrosive inhibitor and pesticides was investigated in samples from both the agricultural runoff and the HTH-PBR effluent. Integrated samples (250 mL) were collected during 4 consecutive days (n=2) and analyzed by gas chromatography coupled to tandem mass spectrometry (GC-MS/MS). Further details on the analytical methodology, including the pre-treatment of the sample, are given elsewhere (Matamoros et al., 2010; Matamoros et al., 2015).

## 3. RESULTS

*3.1. Batch test*

A short batch experiment started with 0.5 m$^3$ of microalgae biomass and 8 m$^3$ of agricultural runoff, as an initial test of the HTH-PBR efficiency. The evolution of the temperature, pH, DO and TSS of the mixed liquor was regularly measured, as well as the alkalinity, COD, $NO_3^-$-N and $PO_4^{-3}$-P concentrations (Table 2). During these days, pH values presented relatively high oscillations due to the photosynthetic activity during the day. Values in the mixed liquor ranged from pH 7.6±0.4 the first day to pH 8.9±0.9 on day 4 (Figure 2). Regarding DO, day and night concentrations oscillated considerably the first day and then remained steady (around 12-14 mg/L). TSS showed a growing trend until the third day of experiment, increasing from 272.5 mg TSS/L on the first day to 325.7 mg TSS/L (Figure 3). Afterwards, values remained rather constant from day 4 (331.8 mg TSS/L). Despite the low nutrient concentrations in the agricultural runoff (0.2 mg N-$NH_4^+$/L, 3.5 mg P-$PO_4^{3-}$/L, 0.1 mg $NO_3^-$-N / L), a relevant growth of the algal biomass was registered, with a 21.7% concentration increase in 4 days.

*3.2. Continuous experiment with agricultural runoff*

Given the positive results obtained in the batch experiment, a long-term performance experiment was carried out, operating the PBR with continuous feeding of agricultural runoff during four months (from 15th December till the end of April) at a hydraulic retention time (HRT) of 16 days (0.5 m$^3$ was replaced per day). Table 3 summarizes the mean values of parameters analyzed daily and/or weekly in both the agricultural runoff and the mixed liquor.

During the winter period (December-February), lower solar radiation (~ 230 W/m$^2$) and temperatures (<10 °C) limited microalgae growth. The nutrients concentration in the water feedstock were also low, with barely 6 mg/L of total Nitrogen and 2.2 mg/L of total Phosphorus. As expected under these nutrient conditions and climate conditions, the biomass concentration of the system barely increased, staying below 240 mg TSS/L until January. A noticeable growth was registered in February, increasing up to 800 mg TSS/L by the end of that month. The corresponding biomass production in this period was in the range of 2-14 g/m$^3$ d. Similar biomass concentrations (approximately 600 mg TSS/L) were obtained in a pilot-scale biofilm reactor in January-February in an airlift tubular PBR located in Cádiz (southern Spain), which worked in continuous mode from December till the end of March (Arbib et al., 2013). In that study, however, the PBR was fed with wastewater effluent from a WWTP nearby, which had similar concentrations of total Phosphorus (around 2.5 mg/L) but higher concentrations of total Nitrogen (around 26 mg/L). Despite of that, TSS values were pretty similar or lower to those obtained in this work. In a different work by Nwoba et al. (Nwoba et al., 2016), they reached a higher productivities in winter (8-47 g /m$^3$ d) using two helical tubular PBRs (40 L each) to treat undiluted digestate from anaerobic digestion piggery effluent, much richer in NH$_4^+$ than the feedstock used in this study. Regarding pH and DO concentrations, both were quite similar to those registered in the agricultural runoff samples, indicating also the low photosynthetic activity during this period.

Not only did the environmental conditions hamper the microalgae growth during the winter period and, in consequence, the treatment efficiency. The low nutrient and organic matter concentrations in the agricultural runoff feeding the HTH-PBR also limited the increase of microalgal biomass production. For instance, sCOD concentrations in the HTH-PBR mixed liquor were higher than the total COD concentrations from the agricultural runoff (average values of 62.6 mg O$_2$-COD/L and 38 mg O$_2$-COD/L, respectively). It has been demonstrated, though, that a fraction of photosynthetically fixed carbon is released during microalgae growth as dissolved organic matter or carbon, and it usually corresponds to 5-30% of the carbon fixed by photosynthesis, but sometimes up to 80%. The higher COD values in the mixed liquor of the PBR could be attributed to this DOM exudation, but also to the low organic matter biodegradability of the agricultural runoff and consequently, to carbon limitation that affected the algal growth. The same behavior was also observed in the work by Arbib et al aforementioned (Arbib et al., 2013). Authors concluded that the low biodegradability of the feedstock could also have led to the consumption of available carbon in the form of carbonate, reducing the alkalinity of the system. A slight decrease in the alkalinity was indeed observed in the HTH-PBR system, as observed in Figure S3.

In spring (March-April), higher solar radiation and temperatures (mean values of 604 W/m$^2$ and 15 °C, respectively) enhanced microalgae photosynthetic activity. These environmental conditions favored the microalgae growth, resulting in higher average pH values (9.7 ±0.5) and DO concentrations (9.2±2.1 mg O$_2$/L) compared to those of the previous period (8.3±1 and 6.9 mg±1.7 O$_2$/L, respectively). Similarly to the winter period, nutrients in the feedstock remained low; however, the N:P ratio increased from 3 to 13 (Table 4) and may have also contributed to the biomass enhancement observed. Indeed, TSS values in the mixed liquor increased from 795 mg L$^{-1}$ in February to 1883 mg L$^{-1}$ by the end of April (see Figure 4). By the end of the spring period, the amount of biomass production reached levels of 76.4 g TSS/m$^3$ d (see Fig. 4b). Under similar working conditions, Arbib et al reached a maximum concentration of 733 mg TSS L$^{-1}$ under an average solar radiation and temperature of 230 W/m$^2$ and 13°C, respectively (Arbib et al., 2013). These results emphasize that the microalgae biomass production seemed to be mainly related to the increase in solar radiation, as nutrient concentrations did not change significantly during the 4 months' performance. Therefore, microalgae biomass production is expected to increase if the environmental conditions were more favorable, regardless of low nutrient and organic matter contents in agricultural runoff. Finally, the performance of the HTH-PBR in the treatment of the agricultural run-off was evaluated during both winter and spring periods, comparing the amounts of nutrients in the agricultural runoff and in the mixed liquor (Table 4). Total Phosphorus was fully removed in the PBR, and total Nitrogen yielded a 95% removal in winter and 84% in spring.

All in all, these preliminary results from both batch and continuous tests reflect the potential of this new design of a full-scale PBR. However, different strategies could be followed in future research works to increase microalgae biomass production, such as reducing the HRT (considering the biofouling that could be derived), adding nutrients to the PBR feedstock (such as digestate from microalgae anaerobic digestion) or CO$_2$ injection. Furthermore, a continuous separation system could be installed to avoid biomass accumulation inside the PBR. The harvested biomass would be further used for co-digestion etc, contributing to make the whole installation more sustainable. All in all, these results can prove very useful to detect the advantages and also the limitations of this system in order to propose different improvements in future PBR designs.

*3.3. Identification and removal efficiency of EOCs*

A sampling campaign was carried out in order to evaluate the removal efficiency of EOCs in the new HTH-PBR system. Four-days integrated water samples from the agricultural runoff and the mixed liquor samples were taken in April, at the end of the continuous experiment. Up to now only a few studies have addressed the capability of microalgae-based systems to remove EOCs, but always referring to open systems (de Godos et al., 2012; Matamoros et al., 2015).

As shown in Figure 5, the highest concentrations detected in the agricultural runoff corresponded to carbamazepine (510 ng/L), benzotriazole (420 ng/L) and tris (2-chloroethyl) phosphate (450 ng/L). The recalcitrant behavior of these three compounds has been reported in numerous studies (Clara , et al., 2004,; Zhang et al., 2008; Loi et al., 2013), which explains these high levels in the runoff water and the poor removal efficiencies registered in the HTH-PBR (<40%). Indeed, neither of these three compounds can be photodegraded, biodegraded or adsorbed, which are the common removal mechanisms taking place during conventional wastewater treatment processes.

The best removal efficiencies corresponded to the anti-inflammatory diclofenac (61%), and the synthetic musk fragrances galaxolide (62%) and tonalide (72%). Similarly to carbamazepine, diclofenac has long been considered a pseudo-persistant contaminant due to the regular background concentration detected in basically all types of environmental waters (Zhang et al., 2008; Gros et al., 2012; García-Galán et al., 2016). Remarkably, elimination rates during wastewater treatment can range from 0% to 90% depending on the study (Gros et al., 2010; Gros et al., 2012). The elimination of diclofenac can be attributed mostly to photodegradation (Zhang et al., 2011) but also to adsorption, as calculated $K_d$ range from 118 to 321 L/Kg (Radjenovic et al., 2009). Actually, Zhang et al. observed the inhibiting effect of the presence of $NO_3^-$-N in water, at concentrations as low as 0.01 mM, during the photodegradation of diclofenac (Zhang et al., 2011). In a recent study by Matamoros et al. in HRAPs, the removal of this anti-inflammatory was 92% during the warm season working at a HRT of 8 days (Matamoros et al., 2015). The removals for ibuprofen and caffeine (<50%) were also low if compared to the 99% removal obtained in the study by Matamoros. Nevertheless, it is complicated to compare both open and closed PBRs in terms of algae cultures and their role in the elimination of contaminants, as the transparency of the HTH-PBR tubes, their roughness and transparency of polyethylene itself are factors to consider (Harris et al., 2013). Biofouling is another challenging problem to overcome when working in PBR systems. In this study, working during winter and early spring helped to avoid this difficult issue.

In the case of the fragrances tonalide and galaxolide, their high volatility and hydrophobicity ($K_{ow}$ of 5.7 and 5.9 respectively) could explain their elimination in the HTH-PBR, enhanced by the high temperatures and the presence of particle matter in the mixed liquor. Concentrations for ibuprofen and caffeine were lower in the agricultural runoff water than values usually found in effluent wastewaters (up to 1.9 µg/L for caffeine and ibuprofen) (Ferrando-Climent et al., 2012). Likewise, elimination efficiencies are usually >90% in conventional WWTPs for both compounds.

4. **Conclusions and perspectives**

This study presents a new design of horizontal tubular hybrid photobioreactor (HTH-PBR) to treat agricultural runoff from drainage channels in rural areas. Likewise, microalgae biomass production is also considered as a highly valuable by-product. Considering a full-scale application in an agricultural context, results from an initial batch test showed an increase in the producitivity of 22.8% (from 272.5 mg TSS/L to 331.8 mg TSS/L) in four days' experiment. During a continuous test, 13.8 g TSS/$m^3$d were obtained in winter, while 74.4 g TSS/$m^3$ d were registered in spring despite the low nutrient concentrations in the feedstock. Removal of different EOCs was also evaluated, with eliminations up to 72% for some fragrances. The relevance of climatic conditions on the overall performance of the systems is highlighted. Future research works will focus in enhancing the efficiency of the reactor, by means of increasing the biomass production and its harvesting.

**Acknowledgements**

This research was funded by the Spanish Ministry of Science and Innovation (Project BIOALGAS, CTM2010-17846). M.J.Garcia and E.Uggetti would like to thank the Spanish Ministry of Industry and Economy for their research grant (FJCI-2014-22767 and IJCI-2014-21594, respectively) and to the European´s Union HORIZON 2020 research and Innovation programme under grant agreement No GA-689242 (INCOVER project). Raquel Gutiérrez would like to thank the Generalitat de Catalunya for her PhD scholarship (2013FI_B 01096). The authors also acknowledge Oscar Durán, Andreu Homs, Clara Latorre and Marc Fructuoso from UPC for their contribution to the experiments, Marionna Hernandez-Mariné from UB for her valuable help on microscopic images.

**FIGURE CAPTIONS**

**Fig. 1.** Flow diagram and sketches of the different parts of the HTH-PBR.

**Fig. 2.** Evolution of pH and dissolved oxygen in the mixed liquor of the HTH-PBR during the batch test with agricultural runoff

**Fig. 3.** Evolution of total suspended solids (TSS) in the HTH-PBR during the batch test

**Fig. 4.** Biomass concentration (a) and biomass production (b) in the HTH-PBR in the long-term experiment. The shadow indicates the first period of experimentation (winter) (n= 24).

**Fig.5.** Removal efficiency of the HTH-PBR working at full capacity under different hydraulic retention times.

**TABLES**

**Table 1.** Technical characteristics of the full-scale hybrid tubular horizontal photobioreactor (HTH-PBR).

| Parameter | Value |
|---|---|
| Type of photobioreactor | Horizontal tubular |
| Total volume (m$^3$) | 8.5 |
| Tank volume (m$^3$) | 0.72 |
| Tank Dimension (m) | 3.6 x 1 |
| Tube volume (m$^3$) | 0.62 |
| Tube diameter (m) | 0.125 |
| Tube length (m) | 50 |
| Number of tubes | 12 |
| Number of tanks | 2 |
| Velocity inside tube (cm/s) | 12.5 |
| Number of engines | 2 |
| Engine power | 750 W |

**Table 2.** Characterization of the feedstock and the HTH-PBR mixed liquor during the batch experiment. ( (n= 32/ n=9 for COD)

| Parameter | Agricultural runoff | Mixed Liquor |
|---|---|---|
| Temperature (ºC) | - | 17.5 (5.6) |
| pH | 8.23 (0.9) | 8.2 (0.8) |
| DO (mg/L) | - | 10.3 (3.5) |
| TSS (mg/L) | - | 299 (26) |
| Alkalinity(mg Ca CO$_3$/L) | 223 | 122 (38) |
| COD(mg/L) | 252 | 140 (20) |
| NO$_3^-$-N(mg/L) | 0.1 | 0.3 |
| PO$_4^{-3}$-P(mg/L) | 3.5 | 3 |

**Table 3**. Characterization of the feedstock and the HTH-PBR mixed liquor during continuous experiment. (n= 24/ n= 10 for COD).

|  | Winter (Dec-Feb 2012) | | Spring (March-April 2013) | |
|---|---|---|---|---|
| **Parameter** | **Agric. Runoff** | **HTH-PBR** | **Agric. Runoff** | **HTH-PBR** |
| **Solar radiation (W/m$^2$)** | 232 (80) | | 604 (212) | |
| **Temperature (ºC)** | 9.4 (2) | | 15 (4) | |
| **pH** | 8.0 (0.3) | 8.3 (1.0) | 8.3 (0.2) | 9.7 (0.5) |
| **DO (mg/L)** | 6.8 (1.4) | 6.9 (1.8) | 8.5 (1.7) | 9.2 (2.2) |
| **Turbidity (NTU)** | 67 (36) | 124 (40) | 74 (55) | 418 (73) |
| **COD (mg/L)** | 38.0 (12)* | 62.6 (31)** | 31 (14)* | 75 (33)** |
| **Alkalinity (mg/L)** | 264 (47) | 96 (29) | 379 (60) | 89 (16) |
| **TS (mg/L)** | 1910 (340) | 1700 (431) | 2496 (400) | 2880 (671) |
| **VS (mg/L)** | 265 (59) | 353 (213) | 316 (74) | 971 (307) |
| **TSS (mg/L)** | 78 (52) | 219 (246) | 59 (19) | 1206 (323) |
| **Biomass Production (g TSS/m$^3$ d)** | - | 4 (1.2) | - | 30 (24) |

Note: average values (SD); DO: Dissolved oxygen  TSS: total suspended solids. TS: total solids. VS: volatile solids. *COD: total chemical oxygen demand. **COD: soluble chemical oxygen demand. All samples were taken at 12 AM.

**Table 4.** Nutrients concentration (mg/L) in the feedstock and mixed liquor of the HTH-PBR during the continuous experiment.

|  | Winter (Dec-Feb 2012) | | Spring (March-April 2013) | |
|---|---|---|---|---|
| **Parameter** | **Agric. Runoff** | **HTH-PBR** | **Agric. Runoff** | **HTH-PBR** |
| **$NH_4^+$-N** | 0.8 (1.5) | 0.3 (0.4) | 4.8 (5) | 0.9 (1.3) |
| **$NO_3^-$-N** | 5 (5) | 0.04 (0.04) | 2.2 (0.3) | - |
| **$NO_2^-$-N** | 0.1 (0.1) | - | 0.5 (0.04) | 0.4 (0.1) |
| **$N_{total}$** | 6 (4.6) | 0.3 (0.2) | 8.9 (6.4) | 1.4 (0.7) |
| **$P-PO_4^{3-}$** | 2.2 (0.7) | - | 0.7 (0.2) | - |
| **$SO_4^{2-}$** | 388 (152) | 311 (35) | 479 (208) | 270 (140) |
| **N:P** | 2,7 | | 12,7 | |
| **$N_{total}$ Removal (%)** | 95,0 | | 84,3 | |

Note: $N_{total}$ comprises $NO_3^-$, $NO_2^-$ and $NH_4^+$.